\documentstyle[11pt,epsfig]{article}
\def\be{\begin{eqnarray}}\def\ba{\begin{eqnarray}}
\def\ee{\end{eqnarray}}\def\ea{\end{eqnarray}}
\def\ben{\begin{enumerate}}\def\bitem{\begin{itemize}}
\def\een{\end{enumerate}}\def\eitem{\end{itemize}}
\def\no{\nonumber\\}

\def\roughly#1{\mathrel{\raise.3ex\hbox{$#1$\kern-.75em%
\lower1ex\hbox{$\sim$}}}}

\def\A0{A_0}
\def\bq{\begin{equation}}
\def\eq{\end{equation}}
\def\la{\langle}\def\ra{\rangle}
\renewcommand{\thefootnote}{\#\arabic{footnote}}
\begin{document}

\begin{titlepage}

\begin{flushright}
\begin{minipage}{3cm}
\begin{flushleft}
DPNU-03-19\\
\today
\end{flushleft}
\end{minipage}
\end{flushright}

\begin{center}
{\Large\bf The Pion Velocity at Chiral Restoration\\
and the Vector Manifestation}
\end{center}
\vspace{1cm plus 0.5cm minus 0.5cm}
\begin{center}
\large Masayasu Harada$^{(a)}$,
Youngman Kim$^{(b)}$, Mannque  Rho$^{(c,d)}$ \\
and Chihiro Sasaki$^{(a)}$
\end{center}
\begin{center}
\end{center}
\vspace{0.5cm plus 0.5cm minus 0.5cm}
\begin{center}
(a)~{\it Department of Physics, Nagoya University, Nagoya,
464-8602, Japan,}\\
(b)~{\it Department of Physics and Astronomy, University of South
Carolina, \\
 Columbia, SC 29208, USA} \\
(c)~{\it Service de Physique Th\'eorique, CEA/DSM/SPhT,
Unit\'e de recherche associ\'ee au CNRS,
CEA/Saclay,  91191 Gif-sur-Yvette c\'edex, France}\\
(d)~{\it School of Physics, Korea Institute for Advanced Study,
Seoul 130-722, Korea}.
\end{center}
\vspace{0.6cm plus 0.5cm minus 0.5cm}

\begin{abstract}
We study the effects of Lorentz non-invariance on the physical
pion velocity at the critical temperature $T_c$ in an effective
theory of hidden local symmetry (HLS) with the ``vector
manifestation" fixed point. We match at a ``matching scale"
$\Lambda_M$ the axial-vector current correlator in the HLS with
the one in the operator product expansion for QCD, and present the
matching condition to determine the bare pion velocity. We find
that the physical pion velocity, which is found to be one at
$T=T_c$ when starting from the Lorentz invariant bare HLS, remains
close to one with the Lorentz non-invariance, $v_\pi (T_c) = 0.83
- 0.99$. This result is quite similar to the pion velocity in
dense matter.

\end{abstract}

\end{titlepage}

\renewcommand{\thefootnote}{\#\arabic{footnote}}
\setcounter{footnote}{0}

\section{Introduction}
Recent developments on effective field theory based on hidden
local symmetry (HLS) suggest a state-of-the-art scenario for the
chiral symmetry restoration at a large $N_f$, high temperature
and/or high density, see Ref.~\cite{hy-review} for a review. In
the ``vector manifestation (VM)" of the HLS theory, the $\rho$
mesons become massless and the longitudinal components of the
$\rho$ mesons and the pions form a chiral multiplet at the
restoration point~\cite{HY:VM}. It has been shown that the vector
manifestation is realized for large $N_f$~\cite{HY:VM}, at the
critical temperature~\cite{HS:T} as well as at the critical
density~\cite{HKR} for chiral symmetry restoration. That the
vector meson mass vanishes at the critical temperature/density
supports the in-medium scaling of the vector meson proposed by
Brown and Rho, ``BR scaling"~\cite{BRscaling}, and has
qualitatively important influences on the properties of hadrons in
medium. It has been shown~\cite{hkrs02} that the HLS with the VM
predicts -- and not posits -- that the vector susceptibility
$\chi_V$ equals the axial-vector susceptibility $\chi_A$ as
required by chiral invariance and that the pion velocity $v_\pi=
1$ with the pion decay constants $f_\pi^t \rightarrow 0$ and
$f_\pi^s\rightarrow 0$ as $T\rightarrow T_c$. This behavior
differs drastically from the scenario predicted by the standard
chiral theory~\cite{SS} where the degrees of freedom at chiral
restoration relevant to the iso-vector susceptibility are the
pions after integrating out the scalar field in the O(4) linear
sigma model
in the case of two-flavor QCD: The standard model prediction is
that the pion velocity goes to zero at the chiral restoration
point.

One of the most important ingredients to realize the VM is the
Wilsonian matching~\cite{HY:WM,hy-review} obtained from the
following general ansatz: Integrating out quark and gluon degrees
of freedom at a matching scale $\Lambda_M$, we obtain the {\it
bare} Lagrangian of the effective field theory (EFT). Then, we can
determine the {\it bare} parameters of the bare Lagrangian by
matching the EFT to the fundamental theory (QCD). 
Physical quantities are obtained by including quantum and thermal and/or
dense effects through the renormalization group equations and
thermal and/or dense loop.
Since we integrate out the high energy modes, i.e., the quarks and
gluons above $\Lambda$, in hot and/or dense matter,
the bare parameters generically have the {\it
intrinsic temperature and/or density dependence}~\cite{HS:T,HKR}
which generally induces Lorentz symmetry violation in the bare
EFT~\cite{hkrs02,hs03}.

In arriving at the HLS/VM results of Ref.~\cite{hkrs02}, it was
assumed that one can ignore Lorentz symmetry breaking in the bare
HLS Lagrangian matched at a scale $\Lambda_M$ to QCD in medium. In
this paper, we lift that assumption. This is made possible by a
``non-renormalization theorem" recently proven by one of the
present authors (C.S.)~\cite{sasaki}. The observation is based on
the existence of a new fixed point in the VM, namely that, {\it
the physical pion velocity does not receive any correction, either
quantum or hadronic thermal, at the critical temperature.} This
means that it suffices to compute the pion velocity at the level
of bare HLS Lagrangian at the matching point to arrive at the
$physical$ pion velocity at the chiral transition. It is important
to study the physical pion velocity since this quantity is a
dynamical object, which controls the pion propagation in medium
through a dispersion relation.

In the present work,  we calculate the $bare$ pion velocity by
matching the axial-vector current correlators given by the HLS
Lagrangian with Lorentz non-invariant terms taken into account, to
the ones from the operator product expansion (OPE) in QCD. In
Section 2, we briefly review HLS to define our notations and write
down the HLS Lagrangian including the Lorentz symmetry violation.
We present in Section 3 the Wilsonian matching condition to
determine the bare pion velocity and its intrinsic temperature
dependence in the low-temperature region. The bare pion velocity
at the critical temperature is calculated in Section 4 by the
Wilsonian matching. Our results are summarized with a brief
conclusion in Section 5.

\section{HLS Theory  Without Lorentz Invariance}
The key observation of hidden local symmetry theory (HLS) is that any
nonlinear sigma model defined in the coset space $G/H$ is
gauge-equivalent to a linear model possessing $G_{\rm
global}\times H_{\rm local}$ symmetry. Here
 $G =
\mbox{SU($N_f$)}_{\rm L} \times \mbox{SU($N_f$)}_{\rm R}$  is the
global chiral symmetry and $H = \mbox{SU($N_f$)}_{\rm V}$ the HLS.
In the HLS theory the basic ingredients are the gauge bosons
$\rho_\mu = \rho_\mu^a T_a$ of the HLS and two SU($N_f$)-matrix
valued variables $\xi_{\rm L}$ and $\xi_{\rm R}$. They are
parameterized as $ \xi_{\rm L,R} = e^{i\sigma/F_\sigma} e^{\mp
i\pi/F_\pi} $, where $\pi = \pi^a T_a$ denote the pseudoscalar
Nambu-Goldstone (NG) bosons associated with the spontaneous
breaking of $G$ and $\sigma = \sigma^a T_a$~\footnote{These
scalars are not to be confused with the scalar that figures in
two-flavor linear sigma model.} the NG bosons absorbed into the
HLS gauge bosons $\rho_\mu$ which are identified with the vector
mesons. $F_\pi$ and $F_\sigma$ are the relevant decay constants,
and the parameter $a$ is defined as $a \equiv F_\sigma^2/F_\pi^2$.
$\xi_{\rm L}$ and $\xi_{\rm R}$ transform as $\xi_{\rm L,R}(x)
\rightarrow h(x) \xi_{\rm L,R}(x) g^{\dag}_{\rm L,R}$, where $h(x)
\in H_{\rm local}$ and $g_{\rm L,R} \in G_{\rm global}$. The
covariant derivatives of $\xi_{\rm L,R}$ are defined by $ D_\mu
\xi_{\rm L} =
\partial_\mu \xi_{\rm L} - i g \rho_\mu \xi_{\rm L}
+ i \xi_{\rm L} {\cal L}_\mu $, and similarly with replacement
${\rm L} \leftrightarrow {\rm R}$, ${\cal L}_\mu \leftrightarrow
{\cal R}_\mu$, where $g$ is the HLS gauge coupling, and ${\cal
L}_\mu$ and ${\cal R}_\mu$ denote the external gauge fields
gauging the $G_{\rm global}$ symmetry. The HLS Lagrangian is given
by
\begin{equation}
{\cal L} = F_\pi^2 \, \mbox{tr} \left[ \hat{\alpha}_{\perp\mu}
\hat{\alpha}_{\perp}^\mu \right] + F_\sigma^2 \, \mbox{tr} \left[
  \hat{\alpha}_{\parallel\mu} \hat{\alpha}_{\parallel}^\mu
\right] + {\cal L}_{\rm kin}(\rho_\mu) \ ,
\label{Lagrangian}
\end{equation}
where ${\cal L}_{\rm kin}(\rho_\mu)$ denotes the kinetic term of
$\rho_\mu$ and
\begin{eqnarray}
&&
\hat{\alpha}_{\perp,\parallel}^\mu =
\left(
  D_\mu \xi_{\rm R} \cdot \xi_{\rm R}^\dag \mp
  D_\mu \xi_{\rm L} \cdot \xi_{\rm L}^\dag
\right) / (2i) \ .
\end{eqnarray}

As we have argued in the Introduction, the application of the
Wilsonian matching in hot matter leads to the intrinsic
temperature dependence of the bare parameters of the HLS
Lagrangian which induces the Lorentz symmetry violation at the
bare level~\cite{hkrs02,hs03}.
The Lorentz non-invariant HLS Lagrangian is constructed
 in Appendix A of Ref.~\cite{HKR}.
At the lowest order, the Lagrangian is given by
\ba
\tilde{\cal L}
&=&\biggl[
  (F_{\pi,{\rm bare}}^t)^2 u_\mu u_\nu
  +
  F_{\pi,{\rm bare}}^t F_{\pi,{\rm bare}}^s
    \left( g_{\mu\nu} - u_\mu u_\nu \right)
\biggr]
\mbox{tr}
\left[
  \hat{\alpha}_\perp^\mu \hat{\alpha}_\perp^\nu
\right]\no
&&+~\biggl[
 (F_{\sigma,{\rm bare}}^t)^2 u_\mu u_\nu
 + F_{\sigma,{\rm bare}}^t F_{\sigma,{\rm bare}}^s
    \left( g_{\mu\nu} - u_\mu u_\nu \right)
\biggr]
\mbox{tr}
\left[
  \hat{\alpha}_\parallel^\mu \hat{\alpha}_\parallel^\nu
\right]
\nonumber\\
&& +~
\Biggl[
  - \frac{1}{ g_{L,{\rm bare}}^2 } \, u_\mu u_\alpha g_{\nu\beta}
  - \frac{1}{ 2 g_{T,{\rm bare}}^2 }
  \left(
    g_{\mu\alpha} g_{\nu\beta}
   - 2 u_\mu u_\alpha g_{\nu\beta}
  \right)
\Biggr]
\,
\mbox{tr}
\left[ V^{\mu\nu} V^{\alpha\beta} \right]
\ ,
\nonumber\\
\label{Lag:no-L}
\end{eqnarray}
where $F_{\pi,{\rm bare}}^t$ ($F_{\sigma,{\rm bare}}^t$) and
$F_{\pi,{\rm bare}}^s$ ($F_{\sigma,{\rm bare}}^s$) denote the {\it
bare} parameters associated with the temporal and spatial decay
constants of the pion (of the $\sigma$). The unit four-vector
needed to account for non-Lorentz-invariance in medium takes the
value $u^\mu=(1,{\bf 0})$ at the rest frame. Here it should be noticed
that, due to the Lorentz symmetry violation, two variables
$\xi_{\rm L}$ and $\xi_{\rm R}$ included in the 1-forms
$\hat{\alpha}_\perp^\mu$ and $\hat{\alpha}_{\parallel}^\mu$ in
Eq.~(\ref{Lag:no-L}) are parameterized as~\cite{sasaki}
\begin{equation}
\xi_{\rm L,R} = e^{i\sigma/F_\sigma^t} e^{\mp i\pi/F_\pi^t}
\ ,
\end{equation}
where $F_{\pi,{\rm bare}}^t$ and
$F_{\sigma,{\rm bare}}^t$
are the
bare parameters associated with the temporal
decay constants of the pion and the $\sigma$.

The terms of ${\cal O}(p^4)$ relevant to the present analysis
are given by
\begin{eqnarray}
\bar{\mathcal L}_{z_2} =
\left[
  2 z_{2,{\rm bare}}^L u_\mu u_\nu g_{\nu\beta}
  + z_{2,{\rm bare}}^T
    \left(
      g_{\mu\alpha}g_{\nu\beta} - 2 u_\mu u_\alpha g_{\nu\beta}
    \right)
\right]
\, \mbox{tr} \left[
    \hat{\mathcal A}^{\mu\nu} \hat{\mathcal A}^{\alpha\beta}
\right]
\ ,
\label{Lag:no-L:z}
\end{eqnarray}
where the parameters $z_{2,{\rm bare}}^L$ and $z_{2,{\rm bare}}^T$
correspond in medium to the vacuum parameter
$z_{2,{\rm bare}}$~\cite{HY:WM,hy-review} at $T=\mu=0$.
$\hat{\mathcal A}^{\mu\nu}$
is defined by
\begin{equation}
\hat{\mathcal A}^{\mu\nu} = \frac{1}{2} \left[
  \xi_{\rm R} {\mathcal R}^{\mu\nu} \xi_{\rm R}^{\dag}
  -
  \xi_{\rm L} {\mathcal L}^{\mu\nu} \xi_{\rm L}^{\dag}
\right]
\ ,
\end{equation}
where ${\mathcal R}^{\mu\nu}$ and ${\mathcal L}^{\mu\nu}$
are the field-strength tensors of the external gauge
fields ${\mathcal R}_\mu$ and ${\mathcal L}_\mu$:
\begin{eqnarray}
{\mathcal R}^{\mu\nu} &=&
  \partial^\mu {\mathcal R}^\nu - \partial^\nu {\mathcal R}^\mu
  - i \left[ {\mathcal R}^\mu \,,\, {\mathcal R}^\nu \right]
\ ,
\nonumber\\
{\mathcal L}^{\mu\nu} &=&
  \partial^\mu {\mathcal L}^\nu - \partial^\nu {\mathcal L}^\mu
  - i \left[ {\mathcal L}^\mu \,,\, {\mathcal L}^\nu \right]
\ .
\end{eqnarray}


\section{Matching Conditions for Bare Pion Velocity}
In this section, we present the matching conditions to determine
the bare pion velocity including the effect of Lorentz symmetry
breaking at the bare level.

The
Wilsonian matching is carried out by matching the vector and
axial-vector current correlators derived from the HLS with those
from the OPE in QCD.
The axial-vector current correlator is defined by
\begin{eqnarray}
G_A^{\mu\nu}(q_0=i\omega_n,\vec{q};T) \delta_{ab}
=
\int_0^{1/T} d \tau \int d^3\vec{x}
e^{-i(\vec{q}\cdot\vec{x}+\omega_n\tau)}
\left\langle
  J_{5a}^\mu(\tau,\vec{x}) J_{5b}^\nu(0,\vec{0})
\right\rangle_\beta
\ ,
\end{eqnarray}
where $J_{5a}^\mu$ is the axial-vector current,
$\omega_n=2n\pi T$ is the Matsubara frequency,
$(a,b)=1,\ldots,N_f^2-1$ denotes the flavor
index and $\langle ~\rangle_\beta$ the thermal average. The
correlator for Minkowski momentum is obtained by the analytic
continuation of $q_0$.
It is convenient to decompose the correlator into
 \ba
G_A^{\mu\nu}(q_0,\vec{q}) 
= q^2 P_L^{\mu\nu} G_A^L(q_0,\bar{q}) +q^2 P_T^{\mu\nu}
G_A^T(q_0,\bar{q}) \ ,\label{decompose}
 \ea
where  we define $\bar{q}=|\vec{q}|$ and
the polarization tensors are given by
 \ba
  P_T^{\mu\nu}
&\equiv&
  g^\mu_i
  \left(
    \delta_{ij} - \frac{q_i q_j}{ \bar{q}^2 }
  \right)
  g_j^\nu\no
&=&
  \left( g^{\mu\alpha} - u^\mu u^\alpha \right)
  \left(
    - g_{\alpha\beta} - \frac{q^\alpha q^\beta}{\bar{q}^2}
  \right)
  \left( g^{\beta\nu} - u^\beta u^\nu \right)
\ ,
\nonumber\\
  P_L^{\mu\nu}
&\equiv&
  - \left( g^{\mu\nu} - \frac{q^\mu q^\nu}{q^2} \right)
  - P_T^{\mu\nu}.
 \ea
The bare Lagrangian is determined at the matching scale $\Lambda$
through the matching and is expressed in terms of the bare parameters
as the sum of Eqs.~(\ref{Lag:no-L}) and (\ref{Lag:no-L:z}).
{}From this bare Lagrangian, the current correlator at the matching
scale is constructed as follows~\footnote{
 The current correlator below $\Lambda$ is obtained by
 including quantum and hadronic corrections through the 
 renormalization group equations and thermal and/or dense loop.
}:
\begin{eqnarray}
&& G_{A{\rm(HLS)}}^L(q_0,\bar{q}) =
  \frac{ F_{\pi,{\rm bare}}^t F_{\pi,{\rm bare}}^s }{
   - [ q_0^2 - V_{\pi,{\rm bare}}^2 \bar{q}^2 ] }
 - 2 z_{2,{\rm bare}}^L
\ ,
\nonumber\\
&&
G_{A{\rm(HLS)}}^T(q_0,\bar{q})
=
-\frac{F_{\pi,\rm bare}^tF_{\pi,\rm bare}^s}{q^2}
  - 2 \frac{ q_0^2 z_{2,{\rm bare}}^L - \bar{q}^2 z_{2,{\rm bare}}^T }
      { q^2 }
\ ,
\label{galt}
\end{eqnarray}
where
$V_{\pi,{\rm bare}}=F_{\pi,{\rm bare}}^s / F_{\pi,{\rm bare}}^t$
is the bare pion velocity.
To perform the matching, we regard $G_A^{L,T}$ as functions of 
$-q^2$ and $\bar{q}^2$ instead of $q_0$ and $\bar{q}$,
and expand $G_A^{L,T}$ in
a Taylor series around $\bar{q}=|\vec{q}|=0$ in $\bar{q}^2/(-q^2)$ as
follows:
\begin{eqnarray}
 G_A^L(-q^2,\bar{q}^2)
 = G_A^{L(0)}(-q^2) + G_A^{L(1)}(-q^2)\bar{q}^2 + \cdots,
\nonumber\\
 G_A^T(-q^2,\bar{q}^2)
 = G_A^{T(0)}(-q^2) + G_A^{T(1)}(-q^2)\bar{q}^2 + \cdots.
\end{eqnarray}
In the following, we determine the bare pion velocity $V_{\pi,{\rm
bare}}$ from $G_A^{L(0)}$ and $G_A^{L(1)}$ via the matching.

Expanding $G_A^{{\rm (HLS)}L}$ in Eq.~(\ref{galt})
in terms of $\bar{q}^2/(-q^2)$, we obtain
\begin{eqnarray}
 &&
 G_A^{{\rm (HLS)}L(0)}(-q^2)
  = \frac{F_{\pi,{\rm bare}}^t F_{\pi,{\rm bare}}^s}{-q^2}
    {}- 2z_{2,{\rm bare}}^L~,
\label{HLS-L(0)}
\\
 &&
 G_A^{{\rm (HLS)}L(1)}(-q^2)
  = \frac{F_{\pi,{\rm bare}}^t F_{\pi,{\rm bare}}^s
    (1 - V_{\pi,{\rm bare}}^2)}{(-q^2)^2}.
\label{HLS-L(1)}
\end{eqnarray}
On the other hand, the correlator $G_A^{\mu\nu}$ in the QCD sector
to be given in OPE is more involved. Our strategy goes as follows.
Since the effect of Lorentz non-invariance in medium has been more
extensively studied in dense matter, we first examine the form of
the relevant correlator in dense matter following
Refs.~\cite{FLK, LM}.
The current correlator $\tilde{G}^{\mu\nu}$ constructed from
the current defined by
\begin{equation}
 J_\mu^{(q)} = \bar{q}\gamma_\mu q, \quad
 \mbox{or} \quad
 J_{5\mu}^{(q)} = \bar{q}\gamma_5\gamma_\mu q,
\end{equation}
is given by
\begin{eqnarray}
 &&
 {\tilde G}^{\mu\nu}(q_0,\bar{q})
 = (q^\mu q^\nu - g^{\mu\nu}q^2)
   \Biggl[ -c_0 \ln |Q^2| + \sum_n \frac{1}{Q^n}A^{n,n} \Biggr]
\nonumber\\
 &&\qquad{}+
   \sum_{\tau = 2}\sum_{k = 1}
   \bigl[ -g^{\mu\nu}q^{\mu_1}q^{\mu_2} + g^{\mu\mu_1}q^\nu q^{\mu_2}
    + q^\mu q^{\mu_1}g^{\nu\mu_2} + g^{\mu\mu_1}g^{\nu\mu_2}Q^2
   \bigr] \nonumber\\
 &&\qquad\qquad \times
   q^{\mu_3}\cdots q^{\mu_{2k}}
   \frac{2^{2k}}{Q^{4k + \tau - 2}}
   A_{\mu_1 \cdots \mu_{2k}}^{2k + \tau, \tau} \nonumber\\
 &&\qquad{}+
  \sum_{\tau = 2}\sum_{k=1}\Bigl[ g^{\mu\nu} - \frac{q^\mu q^\nu}{q^2}
   \Bigr] q^{\mu_1}\cdots q^{\mu_{2k}}
   \frac{2^{2k}}{Q^{4k + \tau -2}}
   C_{\mu_1 \cdots \mu_{2k}}^{2k + \tau, \tau},
\label{acc-ope}
\end{eqnarray}
where $Q^2 = -q^2$. $\tau = d - s$ denotes the twist, and $s = 2k$
is the number of spin indices of the operator of dimension $d$.
Here
$A^{n,n}$ represents the contribution from the Lorentz invariant
operators such as
$A^{4,4} = \frac{1}{6}
  \left\langle \frac{\alpha_s}{\pi} G^2 \right\rangle_\rho$.
$A_{\mu_1 \cdots \mu_{2k}}^{2k + \tau, \tau}$  and $C_{\mu_1
\cdots \mu_{2k}}^{2k + \tau, \tau}$ are the residual Wilson
coefficient times matrix element of dimension $d$ and twist
$\tau$;
e.g., $A_{\mu_1\mu_2\mu_3\mu_4}^{6,2}$ is given by
\begin{equation}
A_{\mu_1\mu_2\mu_3\mu_4}^{6,2} = i
\left\langle
  {\mathcal ST}
  \left(
    \bar{q} \gamma_{\mu_1} D_{\mu_2}D_{\mu_3}D_{\mu_4} q
  \right)
\right\rangle_\rho
\ ,
\end{equation}
where we have introduced the symbol ${\mathcal ST}$ which makes the
operators symmetric and traceless with respect to the Lorentz indices.
The general tensor structure of the matrix element of
$A_{\mu_1 \cdots \mu_{2k}}^{2k + \tau, \tau}$ is given in
Ref.~\cite{hkl}. For $k=2$, it takes the following form:
\ba
&&A_{\alpha\beta\lambda\sigma}=\Bigl[ p_\alpha p_\beta p_\lambda
p_\sigma
    {}- \frac{p^2}{8}\bigl( p_\alpha p_\beta g_{\lambda\sigma}
     {}+ p_\alpha p_\lambda g_{\beta\sigma}
     {}+ p_\alpha p_\sigma g_{\lambda\beta}
     {}+ p_\beta p_\lambda g_{\alpha\sigma}
\nonumber\\
 &&\qquad\qquad
     {}+ p_\beta p_\sigma g_{\alpha\lambda}
     {}+ p_\lambda p_\sigma g_{\alpha\beta} \bigr)
    {}+ \frac{p^4}{48}\bigl( g_{\alpha\beta}g_{\lambda\sigma}
     {}+ g_{\alpha\lambda}g_{\beta\sigma}
     {}+ g_{\alpha\sigma}g_{\beta\lambda} \bigr) \Bigr]A_4
\ea
For $\tau=2$ with arbitrary $k$, we have ~\cite{FLK}:
\ba A_{2k}^{2k+2,2}&=&C_{2,2k}^q
A_{2k}^q+C_{2,2k}^{\rm G} A_{2k}^{\rm G}\no
C_{2k}^{2k+2,2}&=&C_{L,2k}^q A_{2k}^q+C_{L,2k}^{\rm G} A_{2k}^{\rm
G}~,
 \ea
where $C_{2,2k}^q= 1+{\mathcal O}(\alpha_s)$, $C_{L,2k}^{q, {\rm
G}}\sim {\mathcal O}(\alpha_s)$ and $C_{2,2k}^{\rm G}\sim
{\mathcal O}(\alpha_s)$ (with the superscripts $q$ and $G$
standing respectively for quark and gluon) are the Wilson
coefficients in the OPE ~\cite{FLK}. The quantities $ A_{n}^q$ and
$ A_{n}^{\rm G}$ are defined by \ba A_n^q
(\mu)&=&2\int_0^1dx~x^{n-1}[q(x,\mu) + \bar q (x,\mu)]\no A_n^{\rm
G} (\mu)&=&2\int_0^1dx~x^{n-1} G (x,\mu)~, \ea where $q(x,\mu)$
and $G (x,\mu)$ are quark and gluon distribution functions
respectively. We observe that (\ref{acc-ope}) consists of three
classes of terms: One is independent of the background, i.e.,
density in this case, the second consists of scalar operators with
various condensates $\la{\mathcal O}\ra_\rho$ and the third is
made up of non-scalar operators whose matrix elements in dense
matter could not be simply expressed in terms of various
condensates $\la{\mathcal O}\ra_\rho$.

It is clear that Eq.~(\ref{acc-ope}) is a general expression that
can be applied equally well to heat-bath systems. Thus we can
simply transcribe (\ref{acc-ope}) to the temperature case by
replacing the condensates $\la{\mathcal O}\ra_\rho$ by
$\la{\mathcal O }\ra_T$ and the quantities $A_{\mu_1 \cdots
\mu_{2k}}^{2k + \tau, \tau}$ and $C_{\mu_1 \cdots \mu_{2k}}^{2k +
\tau, \tau}$ by the corresponding quantities in heat bath. 
The higher the twist of operators becomes,
the more these operators are suppressed since the dimensions of such
operators become higher and the power of $1/Q^2$ appear.
Thus in the following, 
we restrict ourselves to contributions from the twist 2
$(\tau = 2)$ operators. 
Then the temperature-dependent correlator
can be written as
\begin{eqnarray}
 &&
 G_A^{\mu\nu}(q_0,\bar{q})
 = (q^\mu q^\nu - g^{\mu\nu}q^2)\frac{-1}{4}
   \Biggl[ \frac{1}{2\pi^2}
   \Biggl( 1+\frac{\alpha_s}{\pi} \Biggr) \ln \Biggl( \frac{Q^2}{\mu^2}
   \Biggr) + \frac{1}{6Q^4} \Big\langle \frac{\alpha_s}{\pi} G^2
   \Big\rangle_T \nonumber\\
 &&\qquad\qquad{}- \frac{2\pi\alpha_s}{Q^6} \Big\langle
   \Bigl( \bar{u}\gamma_\mu \gamma_5 \lambda^a u -
   \bar{d}\gamma_\mu \gamma_5 \lambda^a d \Bigr)^2 \Big\rangle_T
\nonumber\\
 &&\qquad\qquad{}- \frac{4\pi\alpha_s}{9Q^6}
   \Big\langle
   \Bigl( \bar{u}\gamma_\mu\lambda^a u + \bar{d}\gamma_\mu\lambda^a d
   \Bigr)\sum_q^{u,d,s}\bar{q}\gamma^\mu\lambda^a q \Big\rangle_T
   \Biggr] \nonumber\\
 &&\qquad{}+
[-g^{\mu\nu}q^{\mu_1}q^{\mu_2} + g^{\mu\mu_1}q^\nu q^{\mu_2}
    + q^\mu q^{\mu_1}g^{\nu\mu_2} + g^{\mu\mu_1}g^{\nu\mu_2}Q^2]\no
&&\qquad\qquad{}\times [\frac{4}{Q^4}A_{\mu_1\mu_2}^{4,2}
+\frac{16}{Q^8}q^{\mu_3}q^{\mu_4}A_{\mu_1\mu_2\mu_3\mu_4}^{6,2}] \
,\label{ope-t} \ea where $G_A^{\mu\nu}$ is constructed from the
axial-vector current associated with the iso-triplet channel
defined by
\begin{eqnarray}
 J_{5\mu}
 &=& \frac{1}{2}(\bar{u}\gamma_5\gamma_\mu u -
     \bar{d}\gamma_5\gamma_\mu d),
\end{eqnarray}
and we keep terms only up to the order of $1/Q^8$ for $ A_{\mu_1
\cdots \mu_{2k}}^{2k+2, 2}$. The $\lambda^a$ denote the $SU(3)$
color matrices normalized as $\mbox{tr}[\lambda^a \lambda^b ] =
2\delta^{ab}$. Here we have dropped the terms with $ C_{\mu_1
\cdots \mu_{2k}}^{2k+2, 2} $ in the non-scalar operators since
they are of higher order in both $1/(Q^2)^n$  and $\alpha_s$
compared to the terms in the first line of Eq. (\ref{ope-t}). The
temperature dependence of $A_{\mu_1\mu_2}^{4,2}$ and
$A_{\mu_1\mu_2\mu_3\mu_4}^{6,2}$, implicit in Eq.~(\ref{ope-t}),
will be specified below.

In order to effectuate the Wilsonian matching, we should in
principle evaluate the condensates and temperature-dependent
matrix elements of the non-scalar operators in Eq.(\ref{ope-t}) at
the given scale $\Lambda_M$ and temperature $T$ in terms of QCD
variables only. This can presumably be done on lattice. However no
complete information is as yet available from model-independent
QCD calculations. We are therefore compelled to resort to indirect
methods and we adopt here an approach borrowed from QCD sum-rule
calculations.

Let us first evaluate the quantities that figure in
Eq.(\ref{ope-t}) at low temperature. In low temperature regime,
only the pions are expected to be thermally excited. In the dilute
pion-gas approximation, $\la {\mathcal O}\ra_T$ is evaluated as
\ba \la {\mathcal O}\ra_T\simeq \la {\mathcal O}\ra_0 +
\sum_{a=1}^3\int\frac{d^3p}{2\epsilon(2\pi)^3}\la\pi^a(\vec p)|
{\mathcal
  O}|\pi^a(\vec p)\ra n_B(\epsilon/T),\label{dpa}
\ea where $\epsilon=\sqrt{\bar p^2+m_\pi^2}$ and $n_B$ is the
Bose-Einstein distribution.
As an
example, we consider the operator of $(\tau,s)=(2,4)$ that
 contributes to both $G_A^{L(0)}$ and $G_A^{L(1)}$.
Noting that $G_A^{L}(q_0,\bar q) =G_{A00}/\bar q^2$, we evaluate
 $G_{A00}(q_0,\bar q)$.
\ba
G_{{\rm A}00}^{(\tau=2,s=4)}(q_0,\bar q)
&=& \frac{3}{4}\int\frac{d^3p}{2\epsilon(2\pi)^3}
\frac{16}{Q^8}[-q^{\alpha}q^{\beta} + g^{0\alpha}q^0 q^{\beta}
    + q^0 q^{\alpha}g^{0\beta} + g^{0\alpha}g^{0\beta}Q^2]\no
&&\qquad\times q^\lambda q^\sigma A_{\alpha\beta\lambda\sigma}^{6,2
 (\pi)}
n_B(\epsilon/T),\label{Ga00} \ea where
$A_{\alpha\beta\lambda\sigma}^{6,2 (\pi)}$ is given
by~\footnote{%
 For the general
 tensor structure of the matrix elements with
 a polarized spin-one target, say, along the beam direction in
 scattering process,
 see Ref. \cite{hjm}.
}
\ba
&&A_{\alpha\beta\lambda\sigma}^{6,2 (\pi)}=\Bigl[ p_\alpha p_\beta p_\lambda p_\sigma
    {}- \frac{p^2}{8}\bigl( p_\alpha p_\beta g_{\lambda\sigma}
     {}+ p_\alpha p_\lambda g_{\beta\sigma}
     {}+ p_\alpha p_\sigma g_{\lambda\beta}
     {}+ p_\beta p_\lambda g_{\alpha\sigma}
\nonumber\\
 &&\qquad\qquad
     {}+ p_\beta p_\sigma g_{\alpha\lambda}
     {}+ p_\lambda p_\sigma g_{\alpha\beta} \bigr)
    {}+ \frac{p^4}{48}\bigl( g_{\alpha\beta}g_{\lambda\sigma}
     {}+ g_{\alpha\lambda}g_{\beta\sigma}
     {}+ g_{\alpha\sigma}g_{\beta\lambda} \bigr) \Bigr]A_4^{\pi}
\ ,\label{A4T}
\ea
where $A_4^\pi$ carries the temperature dependence.
Taking $m_\pi^2=0$, we see that the terms with $p^2$ and
 $p^4$ in Eq. (\ref{A4T}) are zero.

{}From Eqs. (\ref{ope-t}), (\ref{dpa}) and (\ref{Ga00}), we obtain
\begin{eqnarray}
 G_A^{{\rm (OPE)}L(0)}(-q^2)
  &=& \frac{-1}{3}g^{\mu\nu}G_{A,\mu\nu}^{{\rm (OPE)}(0)}
\nonumber\\
  &=& \frac{-1}{4}
   \Biggl[ \frac{1}{2\pi^2}
   \Biggl( 1+\frac{\alpha_s}{\pi} \Biggr) \ln \Biggl( \frac{Q^2}{\mu^2}
   \Biggr) + \frac{1}{6Q^4} \Big\langle \frac{\alpha_s}{\pi} G^2
   \Big\rangle_T
\nonumber\\
 &&{}- \frac{2\pi\alpha_s}{Q^6} \Big\langle
   \Bigl( \bar{u}\gamma_\mu \gamma_5 \lambda^a u -
   \bar{d}\gamma_\mu \gamma_5 \lambda^a d \Bigr)^2 \Big\rangle_T
\nonumber\\
 &&{}- \frac{4\pi\alpha_s}{9Q^6}
   \Big\langle
   \Bigl( \bar{u}\gamma_\mu\lambda^a u + \bar{d}\gamma_\mu\lambda^a d
   \Bigr)\sum_q^{u,d,s}\bar{q}\gamma^\mu\lambda^a q \Big\rangle_T
   \Biggr]
\nonumber\\
&&{}+ \frac{\pi^2}{30}\frac{T^4}{Q^4}A_2^{\pi (u+d)}
  {}- \frac{16\pi^4}{63}\frac{T^6}{Q^6}A_4^{\pi (u+d)}.
\label{OPE-L(0)}
\end{eqnarray}
  $G_A^{{\rm (OPE)}L(1)}$ takes the following form
\ba
 G_A^{{\rm (OPE)}L(1)}
  = \frac{32}{105}\pi^4\frac{T^6}{Q^8} A_4^{\pi (u+d)} .\label{L(1)exp}
 \ea

We now proceed to estimate the pion velocity by matching to
QCD.

First we consider the matching between $G_A^{{\rm (HLS)}L(0)}$ and
$G_A^{{\rm (OPE)}L(0)}$. {}From Eqs.~(\ref{HLS-L(0)}) and
(\ref{OPE-L(0)}), we obtain
\begin{eqnarray}
 (-q^2)\frac{d}{d(-q^2)}G_A^{{\rm (HLS)}L(0)}
 &=& - \frac{F_{\pi,{\rm bare}}^t F_{\pi,{\rm bare}}^s}{Q^2},
\nonumber\\
 (-q^2)\frac{d}{d(-q^2)}G_A^{{\rm (OPE)}L(0)}
 &=& \frac{-1}{8\pi^2}\Biggl[ \Bigl( 1 + \frac{\alpha_s}{\pi} \Bigr) +
   \frac{2\pi^2}{3}\frac{\big\langle \frac{\alpha_s}{\pi}G^2
   \big\rangle_T }{Q^4}
   {}+ \pi^3 \frac{1408}{27}\frac{\alpha_s
    \langle \bar{q}q \rangle_T^2}{Q^6} \Biggr]
\nonumber\\
 &&{}- \frac{\pi^2}{15}\frac{T^4}{Q^4}A_2^{\pi (u+d)}
   {}+ \frac{16\pi^4}{21}\frac{T^6}{Q^6}A_4^{\pi (u+d)}.
\end{eqnarray}
Matching them at $Q^2 = \Lambda_M^2$~\footnote{
 The current correlator is expanded in positive power of $Q^2$ 
 in the HLS theory, while in negative power of $Q^2$ in the OPE.
 Thus we cannot match the whole $Q^2$-dependence between the HLS 
 theory and OPE.
 However if there exists an overlapping area around a scale $\Lambda$,
 we can require the matching condition at that $\Lambda$.
 In fact, the Wilsonian matching at $T=0$ in three flavor QCD was shown
 to give several predictions in remarkable agreement with experiments
 ~\cite{HY:WM,hy-review}.
 This strongly suggests that there exists such an overlapping region.
 As discussed in Ref.~\cite{hy-review}, 
 $\Lambda_M \ll \Lambda_{\rm HLS}$ can be
 justified in the large $N_c$ limit, where $\Lambda_{\rm HLS}$ denotes
 the scale at which the HLS theory breaks down.
 We obtain the matching conditions in $N_c=3$ by extrapolating 
 the conditions in large $N_c$.
 As we mentioned above, the success of the Wilsonian matching at $T=0$
 with taking the matching scale as $\Lambda_M = 1.1\,\mbox{GeV}$
 shows that this extrapolation is valid.  
}, we obtain
\begin{eqnarray}
 \frac{F_{\pi,{\rm bare}}^t F_{\pi,{\rm bare}}^s}{\Lambda_M^2}
 &=& \frac{1}{8\pi^2}\Biggl[ \Bigl( 1 + \frac{\alpha_s}{\pi} \Bigr) +
   \frac{2\pi^2}{3}\frac{\big\langle \frac{\alpha_s}{\pi}G^2
   \big\rangle_T }{\Lambda_M^4}
   {}+ \pi^3 \frac{1408}{27}\frac{\alpha_s
    \langle \bar{q}q \rangle_T^2}{\Lambda_M^6}  \Biggr]
\nonumber\\
 &&{}+ \frac{\pi^2}{15}\frac{T^4}{\Lambda_M^4}A_2^{\pi (u+d)}
   {}- \frac{16\pi^4}{21}\frac{T^6}{\Lambda_M^6}A_4^{\pi (u+d)}
\no
&\equiv& G_0.
\label{ts}
\end{eqnarray}
Next we consider the matching between $G_A^{{\rm (HLS)}L(1)}$ and
$G_A^{{\rm (OPE)}L(1)}$.
{}From Eqs.~(\ref{HLS-L(1)}) and (\ref{L(1)exp}), we have
\begin{equation}
 \frac{F_{\pi,{\rm bare}}^t F_{\pi,{\rm bare}}^s
 (1 - V_{\pi,{\rm bare}}^2)}{\Lambda_M^2}
  = \frac{32}{105}\pi^4\frac{T^6}{\Lambda_M^6} A_4^{\pi (u+d)} .
\label{1-v^2}
\end{equation}
Noting that the right-hand-side of this expression is positive, we
verify that
\begin{equation}
 V_{\pi,{\rm bare}} < 1
\end{equation}
which is consistent with the causality.

The bare pion velocity can be obtained by dividing
Eq.~(\ref{1-v^2}) with Eq.~(\ref{ts}). What we obtain is the
deviation from the speed of light:
\begin{equation}
 \delta_{\rm bare}\equiv 1 - V_{\pi,{\rm bare}}^2
 = \frac{1}{G_0}
   \frac{32}{105}\pi^4\frac{T^6}{\Lambda_M^6} A_4^{\pi (u+d)} .
\label{deviation-rho}
\end{equation}
This should be valid at low temperature. We note that the Lorentz
non-invariance does not appear when we consider the operator with
$s=2$, and that the operator with $s=4$ generates the Lorentz
non-invariance. This is consistent with the fact that $G_A^L$
including up to the operator with $s=2$ is expressed as the
function of only $Q^2$ ~\cite{Mallik:1997kj}.
Equation~(\ref{deviation-rho}) implies that the intrinsic
temperature dependence starts from the ${\cal O}(T^6)$
contribution. On the other hand, the hadronic thermal correction
to the pion velocity starts from the ${\cal O}(T^4)$ ~\cite{hs03}.
[There are ${\mathcal O}(T^2)$ corrections to $[f_\pi^t]^2$ and
$[f_\pi^t f_\pi^s]$, but they are canceled with each other in the
pion velocity.] Thus the hadronic thermal effect is dominant in
low temperature region. At the critical temperature, the
${\mathcal O}(T^2)$ corrections to the pion velocity are also
cancelled as in the low-temperature region. Furthermore there are
no ${\mathcal O}(T^4)$ corrections to either $[f_\pi^t]^2$ or
$[f_\pi^t f_\pi^s]$. Thus hadronic thermal corrections to the pion
velocity are absent due to the protection by the VM. Therefore
there remains only the intrinsic temperature dependence determined
via the Wilsonian matching~\cite{sasaki}.

\section{The Pion Velocity at Critical Temperature}
In this section, we wish to evaluate the physical pion velocity at
the critical temperature $T_c$ starting from the Lorentz
non-invariant bare Lagrangian. According to the
non-renormalization theorem~\cite{sasaki}, the $bare$ velocity so
calculated should correspond to the $physical$ pion velocity $at$
the chiral transition point. Now, in the VM, bare parameters are
determined by matching the HLS to QCD at the matching scale
$\Lambda_M$ and at temperature $T=T_c$.

We begin with a summary of the pion
velocity found in
 the HLS/VM theory with Lorentz invariance~\cite{hkrs02, sasaki, hs03}.
The pion velocity is given
by \cite{hkrs02}\footnote{Note that this definition is equivalent to
  the one used in Ref. \cite{hs03} at one-loop order.}
\begin{eqnarray}
v_\pi^2(\bar{q})
&=&
\frac{
  F_\pi^2(0) +
  \mbox{Re} \, \bar{\Pi}^{s}_\perp (\bar{q},\bar{q};T)
}{
  F_\pi^2(0) +
  \mbox{Re} \, \bar{\Pi}^{t}_\perp (\bar{q},\bar{q};T)
}
\ ,
\label{pv-1}
\end{eqnarray}
where  $\bar{\Pi}^{s,t}_\perp (\bar{q},\bar{q};T)$ is the
axial-vector two-point function~\cite{hkrs02} that represents
hadronic thermal corrections. The two-point function ${\bar
\Pi}_\perp^{\mu\nu}$ is decomposed into four components,
\begin{equation}
{\bar \Pi}_\perp^{\mu\nu}=u^\mu u^\nu {\bar\Pi}_\perp^t +
   (g^{\mu\nu}-u^\mu u^\nu){\bar\Pi}_\perp^s +
   P_L^{\mu\nu}{\bar\Pi}_\perp^L + P_T^{\mu\nu}{\bar \Pi}_\perp^T ~ .
\end{equation}
The VM dictates that if one ignores Lorentz non-invariance in the
bare Lagrangian in medium, the pion velocity approaches the speed
of light as $T\rightarrow T_c$~\cite{hkrs02, hs03}.

In the following, we extend the matching condition valid at low
temperature, Eq.~(\ref{deviation-rho}), to near the critical
temperature, and determine the bare pion velocity at $T_c$. As we
discussed in the previous section, we should in principle evaluate
the matrix elements in terms of QCD variables only in order for
performing the Wilsonian matching, which is as yet unavailable from
model-independent QCD calculations.  Therefore, we make an
estimation by extending the dilute gas approximation adopted in
the QCD sum-rule analysis in the low-temperature region to the
critical temperature with including all the light degrees of
freedom expected in the VM. In the HLS/VM theory, both the
longitudinal and transverse $\rho$ mesons become massless at the
critical temperature since the HLS gauge coupling constant $g$
vanishes. At the critical point, the longitudinal $\rho$ meson
which becomes the NG boson $\sigma$ couples to the vector current
whereas the transverse $\rho$ mesons decouple from the theory
because of the vanishing $g$. Thus we assume that thermal
fluctuations of the system are dominated near $T_c$ not only by
the pions but also by the longitudinal $\rho$ mesons. In
evaluating the thermal matrix elements of the non-scalar operators
in the OPE, we extend the thermal pion gas approximation employed
in Ref.~\cite{hkl} to the longitudinal $\rho$ mesons that figure
in our approach. This is feasible since at the critical
temperature, we expect the equality $A_4^\rho(T_c) = A_4^\pi(T_c)$
to hold as the massless $\rho$ meson is the chiral partner of the
pion in the VM~\footnote{
 We observe from Refs.~\cite{best, SMRS} that
 $A_4^{\pi (u+d)}(\mu=2.4 ~{\rm GeV})\sim  A_4^{\rho
  (u+d)}(\mu=2.4~{\rm GeV})$ even at zero temperature.
}. It should be noted that, although we use the dilute gas
approximation, the treatment here is already beyond the
low-temperature approximation adopted in Eq.~(\ref{dpa}) because
the contribution from $\rho$ meson is negligible in the
low-temperature region. Since we treat the pion as a massless
particle in the present analysis, it is reasonable to take
$A_4^\pi(T) \simeq A_4^\pi(T=0)$. We therefore use
\begin{equation}
 A_4^\rho(T) \simeq A_4^{\pi}(T) \simeq A_4^\pi(T=0)
 \quad \mbox{for}\quad T \simeq T_c.
\label{matrix Tc}
\end{equation}

The properties of the scalar operators giving rise to the
condensates are fairly well understood at chiral restoration. We
know that the quark condensate must be zero at the critical
temperature. Furthermore the value of the gluon condensate at the
phase transition is known from lattice calculations to be roughly
half of the one in the free space~\cite{miller}. We therefore can
use in what follows the following values at $T=T_c$:
 \ba \la\bar q q\ra_T= 0,~~~\la
\frac{\alpha_s}{\pi} G^2\ra_T\sim 0.006 {\rm GeV}^4~.
\label{Tcondensates}
 \ea
Including the contributions from both pions and massless $\rho$
mesons, Eq.~(\ref{L(1)exp}) can be expressed as
 \ba
 G_A^{{\rm (OPE)}L(1)}
  = \frac{32}{105}\pi^4\frac{T^6}{Q^8}\Biggl[ A_4^{\pi (u+d)}+  A_4^{\rho (u+d)} \Biggr].\label{L(1)-pi-rho-exp}
 \ea
Therefore from Eq.~(\ref{deviation-rho}), we obtain the deviation
$\delta_{\rm bare}$ as
\begin{equation}
 \delta_{\rm bare} = 1 - V_{\pi,{\rm bare}}^2
 = \frac{1}{G_0}
   \frac{32}{105}\pi^4\frac{T^6}{\Lambda_M^6}\Biggl[ A_4^{\pi (u+d)}+  A_4^{\rho (u+d)} \Biggr].
\label{deviation-pi-rho}
\end{equation}
This is the matching condition to be used for determining the
value of the bare pion velocity near the critical temperature.

To make a rough estimate of $\delta_{\rm bare}$, we use $ A_4^{\pi
(u+d)}(\mu=1~{\rm GeV}) =0.255$~\cite{hkl}. This value is arrived
at by following Appendix B of ~\cite{hkl}. $A_n^{\pi (q)}$ is
defined by
 \ba \la\pi| \bar q\gamma_{\mu_1} D_{\mu_2}...
D_{\mu_ n}q (\mu)|\pi\ra =(-i)^{n-1} (p_{\mu_1}...p_{\mu_n} -{\rm
traces})A_n^{\pi (q)} (\mu)~, \ea where \ba A_n^{\pi (q)}(\mu)=
2\int_0^1 dx x^{n-1}[q(x,\mu)+(-1)^n \bar q (x,\mu)]~.
 \ea
For any charge state of the pion ($\pi^0, \pi^+, \pi^-$),
$A_n^{\pi (u+d)}$ ($n=2,4$) can be written in terms of the $n$th
moment of valence quark distribution $V_n^\pi (\mu)$ and sea quark
distribution $S_n^\pi (\mu)$~\cite{hkl},
 \ba A_n^{\pi
(u+d)}(\mu)=4V_n^\pi (\mu) +8S_n^\pi (\mu)~, \ea where \ba V_n^\pi
&=&\int_0^1 dx x^{n-1} v^\pi (x,\mu),\no S_n^\pi &=&\int_0^1 dx
x^{n-1} s^\pi (x,\mu).
 \ea
Simple parameterizations of the valence distribution $ v^\pi
(x,\mu)$ and the sea distribution $s^\pi (x,\mu)$ can be found in
Ref. \cite{GRV}-- see Ref.~\cite{GRS} for the updated results --
where the parton distributions in the pions are determined through
the $\pi$-N Drell-Yan and direct photon production processes. With
the leading-order parton distribution functions given in
\cite{GRV}, we obtain $A_4^{\pi (u+d)}=0.255 ~{\rm at}~\mu=1 ~{\rm
GeV}$~\cite{hkl}. For the purpose of comparison with the lattice
QCD result~\cite{best}, we need to calculate the value at $
\mu=2.4 ~{\rm GeV}$; it comes out to be $A_4^{\pi (u+d)}=0.18$.
The value $A_4^{\pi (u+d)}=0.18$ is slightly bigger than the $\sim
0.13$ calculated by the lattice QCD~\cite{best}, while it is a bit
smaller than the $\sim 0.22$ of Ref.~\cite{SMRS}. Note that the
value $\sim  0.18$ agrees with the one determined by lattice QCD
~\cite{best} within quoted errors. Using $A_2^{\pi (u+d)}=0.972$
and $A_4^{\pi (u+d)}=0.255$ and for the range of matching scale
$(\Lambda_M = 0.8 - 1.1\, \mbox{GeV})$, that of QCD scale
$(\Lambda_{QCD} = 0.30 - 0.45\, \mbox{GeV})$ and critical
temperature $(T_c = 0.15 - 0.20\, \mbox{GeV})$, we get
\begin{equation}
 \delta_{\rm bare}(T_c) = 0.0061 - 0.29\,,
\end{equation}
where the $\Lambda_M$ dependence of $A_{2,4}^{\pi (u+d)}$ is
ignored as it is expected to be suppressed by more than
$1/\Lambda_M^6$. Thus we find the $bare$ pion velocity to be close
to the speed of light:
\begin{equation}
 V_{\pi,{\rm bare}}(T_c) = 0.83 - 0.99\,.
\end{equation}
Thanks to the non-renormalization theorem~\cite{sasaki}, i.e., $v_\pi (T_c)=V_{\pi, {\rm bare}}
(T_c)$, we arrive at the physical pion velocity at chiral
restoration:
\begin{equation}
 v_{\pi}(T_c) = 0.83 - 0.99\,.
\end{equation}

\section{Conclusion}

In this paper we lifted the assumption made without justification
in the previous paper~\cite{hkrs02} about the ignorable role of
Lorentz symmetry breaking in the bare Lagrangian which led us to
the conclusion that the pion velocity at the chiral phase
transition equals the speed of light. This feat of doing away with
the assumption was made possible by the non-renormalization
theorem~\cite{sasaki}
that states
that in the HLS/VM theory, the VM protects the pion velocity from
quantum as well as hadronic thermal corrections (at least at one-loop 
level) and hence the $bare$ pion velocity obtained by
matching to QCD at a matching scale $\Lambda_M$ remains
un-renormalized by corrections at the critical temperature. By
using information available from QCD sum rule calculations made in
heat bath, we found that the pion velocity at the chiral
transition temperature remains close to the speed of light. This
result is drastically different from the result obtained in
the standard chiral theory~\cite{SS}
for the
two-flavor QCD which predicts that the pion velocity should go to
zero at the critical point.
The crucial difference between the
HLS/VM result and the
standard chiral theory
result lies in the
degrees of freedom that figure at the phase transition: What
accounts for the drastic difference in the prediction is the
vector mesons becoming massless in the former at the VM fixed
point to which the system is driven by temperature.

It is interesting to compare the result obtained here in heat bath
to a similar result obtained in dense medium, i.e., $v_\pi\sim 1$,
as the system approaches the critical point. In \cite{lprv03}, it
was found that a dense matter is driven to chiral restoration by
the change in the skyrmion background characterized by the
expectation value in ``sliding vacua" (SVEV in short) of the
scalar field that represents the ``soft glue" in QCD trace
anomaly. It turns out that the SVEV of the scalar field relevant
for this process is locked to the quark condensate
$\la\bar{q}q\ra$ in such a way that the melting of the quark
condensate at chiral restoration precisely corresponds to the
melting of the ``soft glue" associated with part of the QCD trace
anomaly. Remarkably the pion velocity is found to approach the
speed of light in dense medium at the critical density in a way
analogous to what is found in this paper for the same quantity at
the critical temperature. 
This suggests that
the vanishing
SVEV of the scalar representing the soft glue in dense medium
might be playing
the role analogous to the VM fixed point in HLS/VM theory
wherein the vector meson mass goes to zero. It remains to be seen
whether the same result is obtained with HLS/VM in dense medium.

\section*{Acknowledgment}

The work of YK is supported  by
the U.S. National Science Foundation,
Grant No. PHY-0140214.


\end{document}